# Improved Approximations for Some Polymer Extension Models

Rafayel Petrosyan[*]

Department of Biosystems Science and Engineering, Eidgenössische Technische Hochschule (ETH) Zurich, 4058 Basel, Switzerland

Department of Physics, University of Alberta, Edmonton, AB T6G2G7, Canada

We propose approximations for force-extension dependencies for the freely jointed chain (FJC) and worm-like chain (WLC) models as well as for extension-force dependence for the WLC model. Proposed expressions show less than 1% relative error in the useful range of the corresponding variables. These results can be applied for fitting force-extension curves obtained in molecular force spectroscopy experiments. Particularly, they can be useful for cases where one has geometries of springs in series and/or in parallel where particular combination of expressions should be used for fitting the data. All approximations have been obtained following the same procedure of determining the asymptotes and then reducing the relative error of that expression by adding an appropriate term obtained from fitting its absolute error.

---

[*] rafayel.petrosyan@alumni.ethz.ch

# Introduction

With the development of single-molecule force spectroscopy (SMFS) in the last 30 years polymer extension models have been extensively used for analyzing force-extension curves [1,2]. One of the most frequently used polymer extension models are freely jointed chain (FJC) and worm-like chain (WLC) models.

The FJC model describes polymer as straight, absolutely rigid segments of an equal length connected by free joints, i.e., the angle between neighboring segments can change without energy penalty. Segments do not have thickness, mass or charge, and are free to pass through each other. Once force is applied on the ends of FJC that extends it, the number of microstates of the system is reduced. The force that resists to the extension of FJC is purely entropic. The exact analytical dependence of extension on force for this model is known to be given by Langevin function: $x(f) = \coth(f) - 1/f$ where $x = z/L$ is the relative extension with $z$ the extension and $L$ the contour length and $f = Fb/k_B T$ is the normalized force with $F$ being the force, $b$ the length of the segment, $k_B$ the Boltzmann constant, and $T$ the absolute temperature [3]. This model is generally used with a single fitting parameter, that is the contour length of the chain – $L$. The length of the segment $b$ then corresponds to the Kuhn length of the real polymer chain. FJC model has been applied for fitting force-extension curves obtained from extending polyinosine in an aqueous medium [4] and polydimethylsiloxane in heptane [5].

The WLC model also known as Kratky-Porod model was first proposed by Kratky and Porod [6,7]. Here, polymer is treated as a long elastic rod without thickness, mass, and charge. As in the case of FJC, here, too, the chain is free to pass through itself. The deformation of a rod can be described with three parameters: bend, stretch, and twist density [6]. In this model, only bending is allowed and it is assumed that there is no twisting or stretching of the chain. Interestingly, the WLC model can also be obtained from FJC model by imposing an additional constraint and then taking the limiting case [8]. First, it is implied that all angles between neighboring segments of FJC are constant and equal, yet segments can rotate freely (so-called freely rotating chain model). Then, WLC can be obtained by going to the limit of infinitely many and infinitely small segments (while the contour length held constant) and by implying that the angles between neighboring segments are tending to be straight (180°). The exact analytical expression for force-extension as well as for extension-force dependencies for the WLC model is not known. However, a popular interpolation formula for force-extension dependence for this model has been invented by Marko and Siggia [9,10]: $f(x) = x - 0.25 + 0.25/(1-x)^2$ where $f = Fp/k_B T$ is the normalized force with $F$ being the force, $p$ the persistence length, $k_B$ the Boltzmann constant, and $T$ the absolute temperature, and $x = z/L$ is the relative extension with $z$ the extension and $L$ the contour length of the chain. WLC model has been extensively applied for fitting force-extension curves of dsDNA and polypeptides in aqueous media using Marko-Siggia approximation [10-12].

As it was mentioned, the FJC and the WLC models assume that polymer does not have thickness, mass, or charge and is free to pass through itself, i.e., different parts of the polymer are free to occupy the same space. To drop this non-physical assumption, the concept of excluded volume was introduced where a potential of interaction between chain components is defined (frequently hard-core potential) [3]. Recently, attempts have been made to find out the force-extension dependence of models where the excluded volume is taken into account [13,14]. Particularly, an interpolation formula has been obtained last year for the WLC model where the hardcore excluded volume effect was assumed 14.

The FJC and WLC phenomenological models are frequently used for fitting force-extension curves with a single fitting parameter – contour length. These continuous models can be used to analyze force-extension curves obtained from polymers with sufficiently large number of repeating units. In order to justify the use of a polymer extension model for the particular type of polymer under the particular external conditions, an experiment can be conducted where the polymer of a known contour length and material properties (such as persistence, Kuhn lengths) can be extended to a certain degree. Then, the model can be fitted to the resulting force-extension data and the fitting parameter(s) can be compared to the known property(s) of the polymer. Depending on the difference between their values, a decision can be made whether the use of that model is appropriate for that polymer under those external conditions.

With the increasing resolution and accuracy of the experimental techniques for performing SMFS [15], more accurate yet simple analytical expressions are required for analyzing force-extension curves. Furthermore, with the possibility of performing more complex SMFS experiments such as when springs are in series [16,17], the knowledge of extension-force dependences of polymer extension models is required. Besides, such analytical expressions can be used for overcoming other problems such as the determination of force dependant loading rate [18], the calculation of the potential energy of the linker system [19], the solution of the Fokker–Planck equation of non-linear dilute polymer [20], and improvement of the algorithm for the automatic alignment of force-extension curves [21]. Thus, in this work, we propose approximation for force-extension dependence for the FJC model and approximations for force-extension as well as extension-force dependencies for the WLC model. We then compare the relative errors of proposed approximations with that of some of the previously obtained approximations. The relative error can be defined as the ratio of the absolute error to the exact value, i.e., $(\text{Exact} - \text{Approximate})/\text{Exact}$ or equivalently $1 - \text{Approximate}/\text{Exact}$. The exact values (evidently, exact to a certain degree) are typically calculated numerically, and the approximate values are provided by the approximation for which the absolute error is being calculated.

## Results and discussion

### Force-extension dependence for the FJC model

As mentioned, the exact analytical extension-force dependence for the FJC model is known: $x(f) = \coth(f) - 1/f$ [3]. The exact analytical force-extension dependence for this model is not known. Several approximations have been proposed and here we will discuss few that are simultaneously simple and accurate. First, in this list is an expression proposed by Puso in his Ph.D. thesis: $f(x) = 3x/(1 - x^3)$ [22]. This expression gives less than 4.62 % of the relative error (Figure 1). Next, a more accurate but more complex expression was proposed by Jedynak: $f(x) = x(3 - 2.6x + 0.7x^2)/(1 - x)(1 + 0.1x)$ [23]. This expression gives less than 1.52 % of the relative error (Figure 1). Soon after, Kröger proposed another expression $f(x) = (3x - 0.2x(6x^2 + x^4 - 2x^6))/(1 - x^2)$ that gives less than 0.28 % of the relative error [24] (Figure 1).

Here, we propose a simple equation for the force-extension dependence Eq. (1) for the FJC model that has less than 0.18 % relative error (Figure 1). The description of the derivation of Eq. (1) is given in the Appendix 1. The exact values for determining the relative error have been calculated using the Langevin function.

$$f(x) = 3x + \frac{x^2}{5}\sin\left(\frac{7x}{2}\right) + \frac{x^3}{1-x} \quad (1)$$

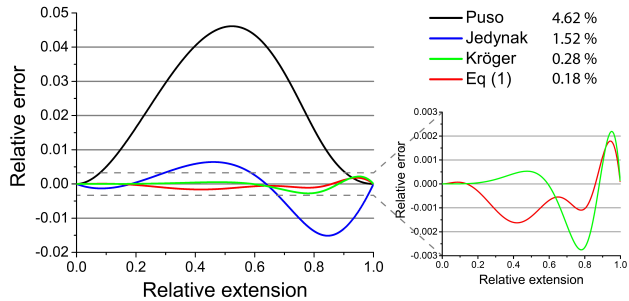

**Figure 1.** Comparison of the relative errors for the approximations for the force-extension dependence for the FJC model. Eq. (1) shows less than 0.18 % relative error. The exact values for determining the relative error have been obtained through the numerical calculation from the Langevin function. The percentages next to the names of the approximations indicate relative error no higher than the given number.

## Force-extension dependence for the WLC model

As mentioned, the exact analytical approximation for the force-extension dependence for the WLC model is not known. Approximations for force-extension dependence for this model have been proposed. The first one was obtained by Marko and Siggia: $f(x) = x - 0.25 + 0.25/(1-x)^2$ [9,10]. This approximation gives less than 17 % of the relative error (Figure 2). Bouchiat, Croquette and colleagues proposed a much more accurate approximation where they have added to Marko-Siggia approximation a 7$^{th}$ order polynomial with coefficients having 5 to 7 decimal digits [25]. In the same work, they provided numerically calculated exact values for the WLC model for the normalized force ranging from 0.1 to 200 [25]. We used those values for calculating the relative errors of expressions discussed. Ogden, Saccomandi and Sgura corrected the Marko-Siggia approximation by adding just a single term $-0.75x^2$ and the resulting expression shows less than 1.9 % of the relative error [26] (Figure 2).

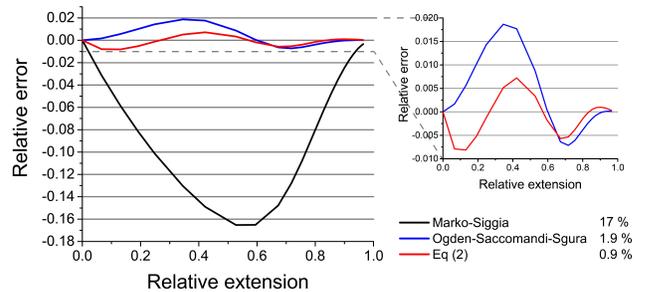

**Figure 2.** Comparison of the relative errors of approximations for the force-extension dependence for the WLC model. Eq. (2) shows less than 0.9 % relative error with respect to the numerical exact values calculated by Bouchiat et al. [25]. The percentages next to the names of the approximations indicate relative error no higher than the given number.

Here we propose a simple approximation for the force-extension dependence for the WLC model Eq. (2) that has less than 0.9 % relative error (Figure 2). The description of the derivation of Eq. (2) is given in the Appendix 2.

$$f(x) = x - 0.8x^{2.15} + \frac{0.25}{(1-x)^2} - 0.25 \quad (2)$$

In order to show the difference between the Eq. (2) and Marko-Siggia approximation on real data, we adopted previously recorded data from the mechanical unfolding of the bacteriorhodopsin [17]. We have focused on the second major unfolding intermediate, since in that case there are relatively many data points that are in the range where the bending elasticity starts to dominate over the entropic elasticity (that is crucial for testing the suitability of the WLC model). As it can be seen in the Figure 3, Eq. (2) shows slightly better fit relative to the Marko-Siggia approximation (this is also seen from the coefficients of determination – $R^2$). The only fitting parameter – contour length, differs by about 0.6 % while all other parameters were fixed to the same values. This small difference in the fitting parameter obtained from fitting of two different approximations for the force-extension behavior of the same model (WLC) justifies the previous use of Marko–Siggia formula for this particular *case*.

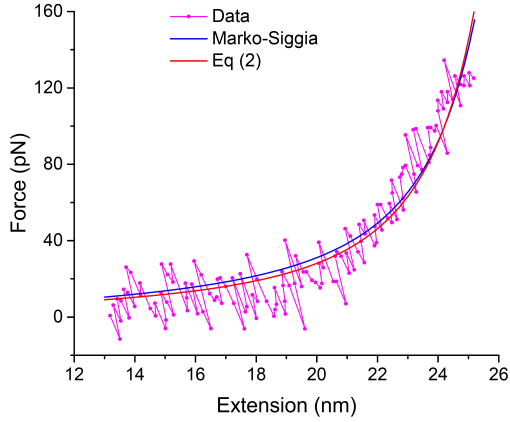

**Figure 3.** Fitting of the Marko-Siggia approximation and Eq. (2) to the second major unfolding intermediate of the bacteriorhodopsin that was adsorbed on mica and mechanically unfolded (the data was adopted from Petrosyan et al. [17]). The second major unfolding intermediate has relatively many data points that are in the range where the bending elasticity starts to dominate over the entropic elasticity (that is crucial for testing the suitability of the WLC model). The contour length obtained through fitting the Marko-Siggia approximation was equal to 28.99973 nm with $R^2$=0.88478. The contour length obtained through fitting the Eq. (2) was equal to 28.83489 nm with $R^2$=0.88804. The persistence length and $k_BT$ were set to 0.4 nm and 4.1 pN nm respectively. Note that both fitted equations represent the same model (WLC) with different accuracies.

## Extension-force dependence for the WLC model

To the best of our knowledge, there were no analytical expressions reported for the extension-force dependencies for the WLC model. Here, we provide an approximation for the extension-force dependence for the WLC model Eq. (3) that has less than 0.95 % relative error with respect to the numerical exact values calculated by Bouchiat et al. [25] (Figure 4). The description of the derivation of Eq. (3) is given in the Appendix 3.

$$x(f) = \frac{4}{3} - \frac{4}{3\sqrt{f+1}} - \frac{10 e^{\sqrt[4]{\frac{900}{f}}}}{\sqrt{f}\left(e^{\sqrt[4]{\frac{900}{f}}} - 1\right)^2} + \frac{f^{1.62}}{3.55 + 3.8 f^{2.2}}$$

(3)

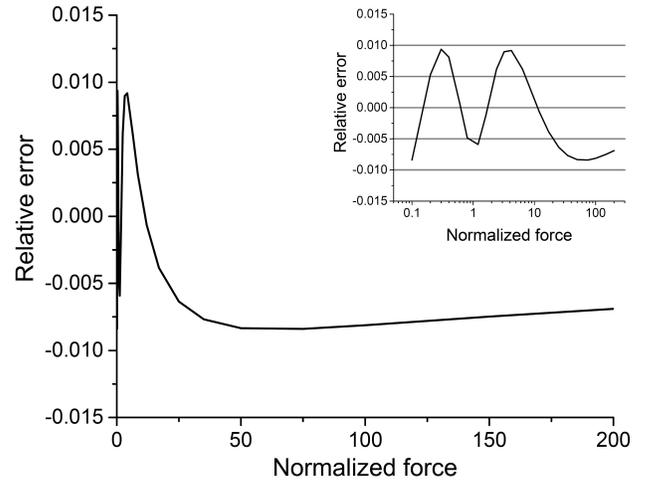

**Figure 4.** The relative error of the approximation for the extension-force dependence for the WLC model. Eq. (3) shows less than 0.95 % relative error in the useful range of the relative extension with respect to the numerical exact values calculated by Bouchiat et al. [25]. The inset shows the relative error of Eq. (3) with respect to the logarithmically scaled normalized force.

## Conclusions

The key results of the current work are three relatively simple and accurate expressions: first one, for the force-extension dependence for the FJC model; second and third ones for the force-extension and extension-force dependencies for the WLC model respectively. We have described the universal method in the Appendix, with which those expressions have been obtained. This heuristic method, probably, could be rigorously formalized in the future.

## Appendix 1

Here, we describe how the Eq. (1) has been obtained. We know the dependence of the relative extension on the normalized force for the FJC model Eq. (4), and we need to determine the dependence of the normalized force on the relative extension for this model.

$$x(f) = \coth(f) - \frac{1}{f} \quad (4)$$

To achieve this goal we adopted an approach that was previously used for deriving approximate formulas [27,28], namely, first determining the asymptotes of the desired dependence, summing those asymptotes and then reducing the relative error of this sum by fitting its absolute error with an appropriate function, and adding the resulting fitting function to that sum of asymptotes.

In order to find the asymptotic behavior for small extensions of the force-extension dependence, we write the Taylor series expansion of the right side of the Eq. (4) for the normalized force tending to 0:

$$x(f) = \frac{f}{3} - \frac{f^3}{45} + O(f^5) \quad (5)$$

Next, by taking only the first term in Eq. (5) for small relative extension, we will have $f(x) \sim 3x$.

For the normalized force tending to infinity from Eq. (4), due to the fact that the exponential terms grow much faster than the linear term, we will have:

$$x(f) = \coth(f) - \frac{1}{f} = \frac{e^{2f}+1}{e^{2f}-1} - \frac{1}{f} \sim 1 - \frac{1}{f} \quad (6)$$

Hence, for this limiting case, we have $f(x) \sim 1/(1-x)$. Next, we need to sum up these two limiting cases. However, care should be taken so that these asymptotes will not interfere with each other. For example, the term $1/(1-x)$ equals to 1 when the relative extension is 0; this means that this term should be multiplied with some function that will make it 0 when the relative extension is 0 and it will still go to infinity as $1/(1-x)$ when the relative extension tends to 1; clearly, the multiplier can be higher than the first power of the relative extension so that term $3x$ dominates when $x$ tends to 0. We found that the relative error was minimal among the multipliers with an integer powers of $x$ when this power was equal to 3. Now we have the following approximation:

$$f(x) = 3x + x^3/(1-x) \quad (7)$$

Interestingly, Eq. (7) is identical to the expression proposed by Darabi and Itskov $f(x) = x(x^2 - 3x + x)/(1-x)$ [29] which is identical to the expression found in the work by Gou, Ray and Akhremitchev $f(x) = 1/(1-x) - (1-x)^2$ [30]. Next the relative error of Eq (7) was minimized by fitting its absolute error with the appropriate function and then adding that function with the resulting fitting parameters to the Eq. (7). As earlier here, too, care should be taken so that the fitting function will not interfere with the asymptotes obtained previously.

In this case, we have found that the simple function of the form $ax^2\sin(bx)$ is fitting well the absolute error of Eq. (7). Last error correction step can be repeated by fitting the absolute error of the last approximation with an appropriate function and then again adding it to that approximation. This high accuracy will be at the cost of the simplicity of the expression. Such accurate approximations, with more than one correction steps, have been obtained for the invers Langevin function previously [31,32].

## Appendix 2

The obtainment of the Eq. (2) was relatively simpler, since in this case, the asymptotes have been already determined by Marko and Siggia [9] $f(x) = x - 0.25 + 0.25/(1-x)^2$ and only the last step, namely, the determination of the appropriate function for fitting the absolute error of the Marko-Siggia approximation was required. In this case, we have found that the simple function of the form $ax^b$ is fitting well the absolute error of the Marko-Siggia approximation. Then, that term has been added to the Marko-Siggia approximation and the resulting approximation had significantly smaller relative error (Figure 2).

## Appendix 3

The derivation of the Eq. (3) was not that straightforward, since the determination of the asymptotes that are not interfering with each other was not trivial for this case. Note that when the relative extension tends to 0, the Taylor series expansion of the Marko-Siggia expression will give the following:

$$f(x) = \frac{3x}{2} + \frac{3x^2}{4} + O(x^3) \quad (8)$$

By taking only the first term in Eq. (8) for small normalized forces, we will have $x(f) \sim 2f/3$.

When the relative extension tends to 1, the normalized force tends to infinity as $f(x) \sim 1/4(1-x)^2$. Hence, for this limiting case we have $x(f) \sim 1 - 1/2\sqrt{f}$.

Let us understand why the Eq. (9) has been chosen for building the final approximation – Eq. (3).

$$x(f) = \frac{4}{3} - \frac{4}{3\sqrt{f+1}} - \frac{10e^{\sqrt[4]{\frac{900}{f}}}}{\sqrt{f}\left(e^{\sqrt[4]{\frac{900}{f}}} - 1\right)^2} \quad (9)$$

When the normalized force tends to 0, the Taylor series expansion of the first two summands of the Eq. (9) will be given by Eq. (10)

$$\frac{4}{3} - \frac{4}{3\sqrt{f+1}} = \frac{2f}{3} - \frac{f^2}{2} + O(f^3) \quad (10)$$

Note that when the normalized force tends to 0, the third summand of Eq. (9) tends to 0 exponentially, i.e., much faster than the linear term $2f/3$ and does not interfere with it.

When the normalized force tends to infinity, the Taylor series expansions of the first two summands and the third summand of Eq. (9) will be given by Eqs. (11) and (12) respectively.

$$\frac{4}{3} - \frac{4}{3\sqrt{f+1}} = \frac{4}{3} - \frac{4}{3\sqrt{f}} + O\left(\left(\frac{1}{f}\right)^{\frac{3}{2}}\right) \quad (11)$$

$$\frac{10e^{\sqrt[4]{\frac{900}{f}}}}{\sqrt{f}\left(e^{\sqrt[4]{\frac{900}{f}}}-1\right)^2} = \frac{1}{3} - \frac{5}{6\sqrt{f}} + O\left(\frac{1}{f}\right) \quad (12)$$

By considering only the first two terms on the right sides in Eqs. (11) and (12) and subtracting the Eq. (12) from the Eq. (11), we can obtain the desired limiting behavior for the relative extension when the normalized force tends to infinity. The Eq. (9) can be further improved by fitting its absolute error with the function of the following form: $f^a/(b+cf^d)$ and then adding this function with the resulting fitting parameters to the Eq. (9). Thus Eq. (3) can be obtained.